\documentclass[twocolumn,showpacs,preprintnumbers,amsmath,amssymb]{revtex4}
 \usepackage{graphicx}  

\newcommand{\beq}{\begin{eqnarray}}
\newcommand{\eeq}{\end{eqnarray}}

\begin{document}

\title{Hyperon mixing and universal many-body repulsion in neutron stars}

\author{Y.\ Yamamoto$^{1}$}
\author{T.\ Furumoto$^{2}$}
\author{N.\ Yasutake$^{3}$}
\author{Th.A.\ Rijken$^{4}$$^{1}$}
\affiliation{
$^{1}$Nishina Center for Accelerator-Based Science,
Institute for Physical and Chemical
Research (RIKEN), Wako, Saitama, 351-0198, Japan\\
$^{2}$National Institute of Technology, Ichinoseki College, Ichinoseki, 
Iwate, 021-8511, Japan\\
$^{3}$Department of Physics, Chiba Institute of Technology, 2-1-1 Shibazono
Narashino, Chiba 275-0023, Japan\\
$^{4}$IMAPP, University of Nijmegen, Nijmegen, The Netherlands
}

%

\begin{abstract}
A multi-pomeron exchange potential (MPP) is proposed as a model for
the universal many-body repulsion in baryonic systems on the basis 
of the Extended Soft Core (ESC) bryon-baryon interaction. 
The strength of MPP is determined by analyzing the nucleus-nucleus 
scattering with the G-matrix folding model. 
The interaction in $\Lambda N$ channels is shown to reproduce well
the experimental $\Lambda$ binding energies.
The equation of state (EoS) in neutron matter with hyperon mixing is 
obtained including the MPP contribution, and mass-radius relations of 
neutron stars are derived.
It is shown that the maximum mass can be larger than the observed one 
$2M_{\odot}$ even in the case of including hyperon mixing
on the basis of model-parameters determined by terrestrial experiments.
\end{abstract}


\maketitle

\parindent 15 pt

\section{Introduction}

It is a fundamental problem to understand properties of
baryonic many-body systems such as nuclei, hypernuclei 
and neutron stars on the basis of underlying 
baryon-baryon (BB) interactions.
The basic property of nuclear systems composed of nucleons
is the saturation property of density and energy per particle.
Though important roles for this property are played by repulsive 
cores and tensor components included in nucleon-nucleon ($NN$)
interactions, it is insufficient quantitatively:
We need to take into account the three-nucleon interaction
composed of the attractive part (TNA) and the repulsive part (TNR).
While the TNA contributes moderately as function of density, 
the TNR contribution increases rapidly in the high-density region 
and leads to high values of the nuclear incompressibility.
It is well known that the latter is indispensable for
the stiff equation of state (EoS) of neutron-star matter
needed to reproduce large maximum masses of neutron stars.

The neutron stars J1614-2230~\cite{Demorest10} and
J0348-0432~\cite{Antoniadis13} have brought great impacts on 
the maximum-mass problem, observed masses of which are 
$(1.97\pm0.04)M_{\odot}$ and $(2.01\pm0.04)M_{\odot}$, respectively.
These large masses give a severe condition for the stiffness of
EoS of neutron-star matter, suggesting the existence of strong TNR.

On the other hand, the hyperon ($Y$) mixing in neutron-star matter 
is known to bring about the remarkable softening of the EoS, which 
cancels the TNR effect for the maximum mass~\cite{Baldo00,Vidana00,NYT}. 
One of ideas to avoid this serious problem is to consider that
the TNR-like repulsions work universally for $Y\!N\!N$, $Y\!Y\!N$ 
$Y\!Y\!Y$ as well as for $N\!N\!N$ \cite{NYT}.
In this work we adopt this idea: Universal repulsions among
three baryons are called here as the three-baryon repulsion (TBR).
The main subject in this paper is to investigate whether or not the
maximum mass of $2M_{\odot}$ can be obtained from the EoS for 
hyperon-mixed neutron-star matter, when TBR is taken into account.

In order to treat hyperon-mixed nuclear matter realistically,
it is indispensable to use a reliable interaction model for 
baryon-baryon ($B\!B$) channels including not only $N\!N$ but also
$Y\!N$ and $Y\!Y$: We adopt here the Extended Soft Core (ESC) model 
developed by two of authors (T.R. and Y.Y.) and M.M. Nagels~\cite{ESC08}. 
In this model two-meson and meson-pair exchanges are taken into 
account explicitly and no effective boson is included differently 
from the usual one-boson exchange models. The latest version of 
ESC model is named as ESC08c~\cite{ESC08,ESC08c}.
Hereafter, ESC means this version.
The TBR is taken into account by the 
multi-pomeron exchange potential (MPP) within the ESC modeling.


Some many-body theory is needed to treat many-body systems
with a realistic $B\!B$ interaction model:
The G-matrix theory is a good tool for such a purpose,
where the correlations induced by short-range and tensor components
are renormalized into G-matrix interactions.
Similarly baryonic coupling terms such as $\Lambda\!N$-$\Sigma\!N$ ones
are included this way into single-channel G-matrices such as 
$\Lambda\!N$-$\Lambda\!N$ ones.
In the case of nucleon matter, the lowest-order G-matrix calculations 
with the continuous (CON) choice for intermediate single particle potentials
were shown to simulate well the results including higher hole-line contributions
up to about 4 times of normal density $\rho_0$~\cite{Baldo02}.
On the basis of this recognition, we study properties of baryonic matter 
including not only nucleons but also hyperons 
with use of the lowest-order G-matrix theory with the CON choice.

The methodology in our works is to use the $B\!B$ interaction model
determined on the basis of terrestrial experiments,
namely to introduce no ad hoc parameter to stiffen the EoS.
The most important is how to determine the strength of the TNR 
being an essential quantity for the stiffness of EoS.
Many attempts have been made to extract some information on the 
incompressibility $K$ of high-density matter formed in high-energy 
central heavy-ion collisions. In many cases, however,
the results for the EoS still remain inconclusive.
On the other hand, it was shown clearly in ref.\cite{FSY}
that the TNR effect appeared in angular distributions of 
$^{16}$O+$^{16}$O elastic scattering ($E/A$=70 MeV), $etc$.
Such a scattering phenomenon can be analyzed quite successfully
with the complex G-matrix folding potentials derived from
free-space $N\!N$ interactions.
Then, the G-matrix folding potentials including MPP contributions
are used to analyze the $^{16}$O+$^{16}$O scattering, and 
the strengths of MPP are adjusted so as to reproduce the experimental data.
The determined MPP interactions are included in constructing the EoS of 
neutron-star matter, being expected to result in a stiff EoS enough to give
the observed neutron-star mass~\cite{YFYR}. 
It should be noted that our MPP is defined so as to work universally 
not only in $N\!N\!N$ states but also $Y\!N\!N$, $Y\!Y\!N$ and $Y\!Y\!Y\!$ states.
Corresponding to the determined MPP, the TNA is added phenomenologically
to reproduce the nuclear saturation property precisely.


Thus, our $B\!B$ interaction is composed of ESC, MPP and TNA.
ESC gives potentials in $S=-1$ ($\Lambda\!N$, $\Sigma\!N$) and 
$S=-2$ ($\Xi\!N$, $\Lambda\!\Lambda$ and $\Lambda\!\Sigma$) channels. 
MPP is universal in these channels.
Because TNA is given in $N\!N$ channels phenomenologically, 
there is no theoretical correspondence in $S<0$ channels.
However, we can confirm the validity of ESC+MPP+TNA model
in these channels by applying this interaction to hypernuclear calculations,
and then TNA is considered as a three-baryon attraction (TBA).

The final step in this work is to study properties of neutron stars
with hyperon mixing on the basis of our $B\!B$ interaction model.
The EoS of $\beta$-stable neutron-star matter composed of
neutrons ($n$), protons ($p^+$), electrons ($e^-$), muons ($\mu^-$)
and hyperons ($\Lambda$ and $\Sigma^-$) is derived from the G-matrix 
calculation with use of the ESC+MPP+TBA model.
Using the EoS of hyperonic neuron-star matter, we solve the 
Tolmann-Oppenheimer-Volkoff (TOV) equation for the hydrostatic structure,
and obtain mass-radius relations of neutron stars. 

For a massive neutron star including hyprons, there are the works 
based on the relativistic mean field models \cite{Weiss,Bednarek,Jiang}.
In comparison with these works, the feature of our approach is to
start from the well-established $B\!B$ interaction model, and to use no 
adjustable parameter except those in the additional many-body interactions
determined in terrestrial experiments.

This paper is organized as follows:
In Sect.II, $B\!B$ interaction models are introduced.
It is explained how to determine MPP and TBA parts.
In Sect.III, ESC+MPP+TBA model is tested by comparing the 
calculated result for $\Lambda$ hypernuclei to experimental data.
In Sect.IV, we derive the EoS of hyperonic nuclear matter.
By solving the TOV equation, the mass-radius relations are obtained.
The conclusion of this paper is given in Sect.V.

\section{Interaction model}

\subsection{Baryon-Baryon interaction ESC}

In Nijmegen ESC-potentials, all available $N\!N$-, $Y\!N$-, and $Y\!Y$-data are 
fitted simultaneously with single sets of meson parameters.
In the most recently developed ESC-model (ESC08c) 
the dynamics consists of the following ingredients:
\begin{enumerate}
\item[(i)] OBE potentials from 
pseudo-scalar ($J^{PC}=0^{-+}$), vector ($J^{PC}=1^{--}$), scalar ($J^{PC}=0^{++}$)
and axial-vector ($J^{PC}=1^{++}$) are treated with the most general vertices.
Besides these, also included are the axial-vector mesons with $J^{PC}=1^{+-}$.
Two-meson-exchange (TME) potentials in ESC are restricted to 
two-pseudoscalar-exchange (ps-ps) potentials, where the full pseudoscalar nonets
are exchanged. 
\item [(ii)]  
Meson-pair exchange (MPE). 
The two-meson-baryon-baryon vertices are the low energy
approximations of (a) the heavy-meson and their two-meson decays, and (b) 
contributions from baryon-resonance $\Delta_{33}$ etc. and negative-energy states.
The MPE-interactions have been extended to all $\{8\}\otimes\{8\}$ BB-channels by
using SU$_f$(3)-symmetry. For example the Tomozawa-Weinberg pair-interaction
potential is included in ESC.
\item [(iii)]  
Diffractive contributions to the soft-core potential. 
The pomeron is thought of being related to an even number of gluon-exchanges. 
Next to the Pomeron-exchange (even number of gluons) also
Odderon-exchange (odd number of gluons) is included in the OBE-part of the interactions.
Also, room is made for quark-core effects supplying extra repulsion,
which may be required in some $BB$-channels
such as $\Sigma^+ p(I=3/2,^3S_1)$- and $\Sigma N(I=1/2,^1S_0)$-channels. 
We describe this structural effect phenomenologically
by Gaussian repulsions, similar to the pomeron. In ESC the strength of 
this repulsion is taken proportional to the weights of the 
SU(6)-forbidden $[51]$-configuration in the various $B\!B$-channels.
\end{enumerate}


As a model of universal TBR, we introduce
the multi-pomeron exchange potential (MPP) \cite{ESC08},
consistently with the ESC modeling, assuming that
the dominant mechanism is triple and quartic pomeron exchange.


The three- and four-body local potentials are derived from
the triple- and quartic-pomeron vertexes.
The density($\rho$)-dependent two-body potentials in a baryonic medium are 
obtained by integrating over coordinates of third (and fourth)
particles in the three-body (and four-body) potentials as follows:
\begin{eqnarray}
V^{(3)}_{eff}(r) 
&=& g_P^{(3)} (g_P)^3 \frac{\rho}{{\cal M}^5} F(r) \ ,
\\
V^{(4)}_{eff}(r) 
&=& g_P^{(4)} (g_P)^4 \frac{\rho^2}{{\cal M}^8} F(r) \ ,
\\
F(r) &=& \frac{1}{4\pi} \frac{4}{\sqrt{\pi}}
\left(\frac{m_P}{\sqrt{2}}\right)^3
\exp\left(-\frac12 m_P^2 r^2 \right) \ .
\label{eq:tbf.3}
\end{eqnarray}
Here, the values of the two-body pomeron strength $g_P$ and 
the pomeron mass $m_P$ are the same as those in ESC.
A scale mass ${\cal M}$ is taken as a proton mass.




\subsection{Determination of MPP strength}


%
In the same way as \cite{YFYR},
the analyses for the $^{16}$O$+^{16}$O elastic scattering at 
an incident energy per nucleon $E_{in}/A=70$ MeV
are performed so that the MPP strengths $g_P^{(3)}$ and $g_P^{(4)}$ are 
determined to reproduce the experimental data with the use of the G-matrix 
folding potential derived from ESC including MPP.

Because the nuclear saturation property cannot be reproduced only 
by adding MPP to ESC, we introduce also an attractive part phenomenologically 
as a density-dependent two-body interaction
\begin{eqnarray}
V_{A}(r;\rho)= V_{0}\, \exp(-(r/2.0)^2)\, \rho\, 
\exp(-\eta \rho)\, (1+P_r)/2 \ ,
\end{eqnarray}
$P_r$ being a space-exchange operator. 
Here, because the functional form is not determined within 
our analysis, it is fixed to be similar to the TNA part 
given in \cite{Panda81}. $V_{0}$ and $\eta$ are 
treated as adjustable parameters.
$V_{A}(r;\rho)$ works only in even states due to a $(1+P_r)$ factor.
This assumption is needed to reproduce the $^{16}$O$+^{16}$O potential
at $E/A=70$ MeV and nuclear-matter energy consistently \cite{YFYR}.

On the basis of G-matrix calculations, strengths of
the MPP part ($g_P^{(3)}$ and $g_P^{(4)}$) and the attractive part 
($V_0$ and $\eta$) are determined so as to reproduce the 
$^{16}$O$+^{16}$O angular distribution at $E_{in}/A=70$ MeV, 
and to reproduce the saturation properties of nucleon matter.
The determined parameters are listed in Table \ref{MPP}.
\begin{table}[ht]
\centering 
\setlength{\textwidth}{50mm} 
\caption{Parameter values included in MPP and TNA.}
\label{MPP}
\vskip 0.2cm
 \begin{tabular}{ccccc}
 \hline\hline
& $g_P^{(3)}$ & $g_P^{(4)}$ & $V_0$ & $\eta$ \\ 
 \hline
 (a)& 2.34 & 30.0  & $-$32.8 &  3.5  \\
 (b)& 2.94 & 0.0   & $-$45.0 &  5.4  \\
 (c)& 2.34 & 0.0   & $-$43.0 &  7.3  \\
\hline\hline 
\end{tabular}
\end{table}
Here, it should be noted that the ratio of $g_P^{(3)}$ and $g_P^{(4)}$ 
cannot be determined in our analysis.
In the same way as Ref.\cite{YFYR}, we choose it rather adequately
referring the estimation given in \cite{Kai74,Bron77}.
Then, chosen values of $g_P^{(3)}$ and $g_P^{(4)}$ are included
in set (a). 
On the other hand, $g_P^{(4)}$= 0 is taken in sets (b) and (c).
Set (b) is determined to reproduce $^{16}$O$+^{16}$O angular distribution
as well as set (a). Set (c) has the same value of $g_P^{(3)}$ as set (a).
Hereafter, interactions ESC+MPP+TNA with sets (a), (b) and (c) are
named MPa, MPb and MPc, respectively.

The basic properties of nucleon matter are given by
the following quantities:
Denoting an energy per particle as $E(\rho,\beta)$
with $\beta=(\rho_n-\rho_p)/\rho$,
a symmetric energy $E_{sym}$ and its slope parameter $L$
are expressed as
$E_{sym}=\frac12 \left[ \frac{\partial^2 
E(\rho,\beta)}{\partial \beta^2}\right]_{\rho_0}$ and
$L=3\rho_0 \left[\frac{\partial E_{sym}(\rho)}{\partial \rho}
\right]_{\rho_0}$, respectively.
A incompressibility of symmetric nucleon matter is give by
$K=9\rho_{\rho_0}^2\left[\frac{\partial^2 
E(\rho,0)}{\partial \rho^2}\right]_{\rho_0}$.

The $E/A$ values for MPa/b (MPc) are $-15.8$ ($-15.5$) MeV
at the saturation density $\rho_0$=0.16 fm$^{-3}$.
The values of $E_{sym}$ at $\rho_0$ are 33.1, 33.1 and 32.7 MeV
in the cases of MPa, MPb and MPc, respectively, and
the values of $L$ are 70, 69 and 67 MeV correspondingly.
These values are in nice agreement to the values 
$E_{sym}=32.5\pm0.5$ MeV and $L=70\pm15$ MeV determined 
recently on the basis of experimental data~\cite{Yoshida}.
The obtained values of $K$ at $\rho_0$ are 310, 280 and 260 MeV
for MPa/b/c, respectively.
Thus, the nuclear saturation property derived from MPa/b/c
is quite reasonable in comparison with the empirical values.

The MPP parts in MPa and MPb are the same as MP1a and MP2a
given in Ref.\cite{YFYR}, respectively.
The differences are in the TNA parts:
Those of MPa and MPb are tuned so as to reproduce
the saturation properties more accurately than those of
MP1a and MP2a.

\begin{figure}[htb!] 
\centering 
\includegraphics[width=8cm]{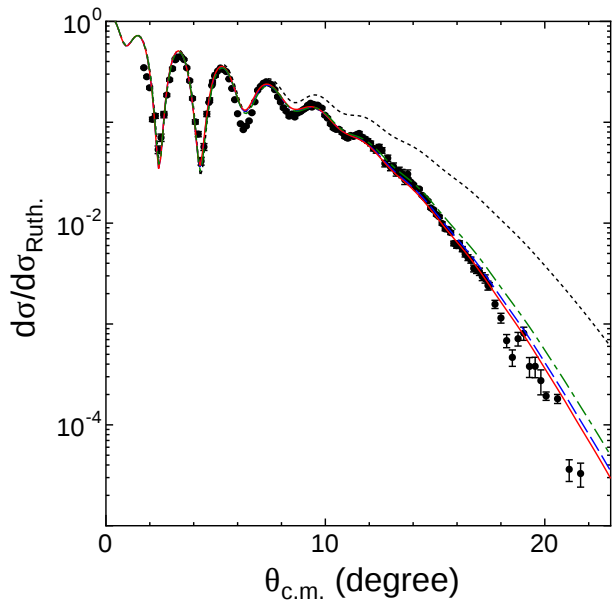} 
\caption{\small \label{xsO16O16}
(Color online) Differential cross sections for $^{16}$O+$^{16}$O elastic 
scattering at $E/A=70$ MeV calculated with the G-matrix folding potentials.
Solid, dashed and dot-dashed curves are for MPa, MPb and MPc, respectively.
Dotted curve is for ESC. 
}
\end{figure} 

\begin{figure}[htb!] 
\centering 
\includegraphics[width=8cm]{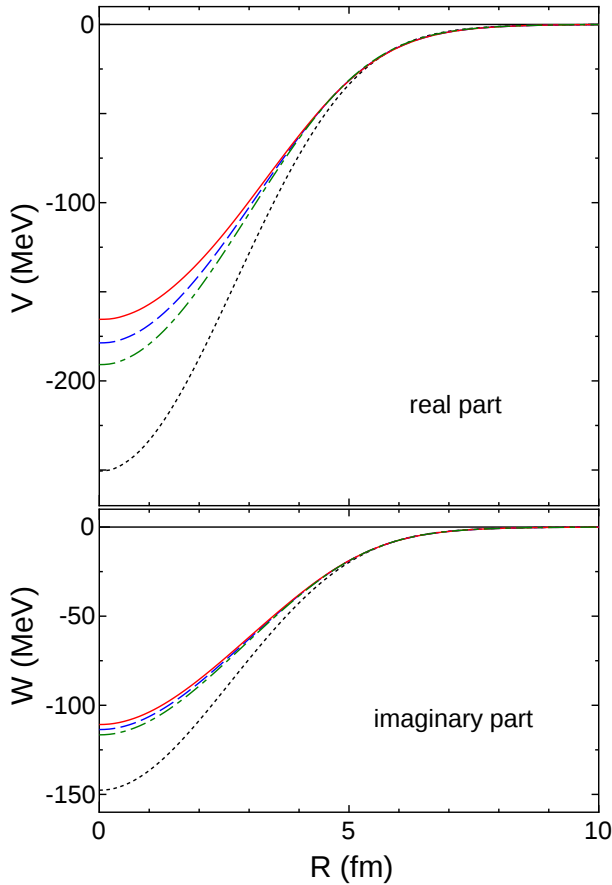} 
\caption{\small \label{potO16O16}
(Color online) Double-folding potentials for $^{16}$O+$^{16}$O elastic 
scattering at $E/A=70$ MeV. Solid, dashed and dot-dashed curves are for
MPa, MPb and MPc, respectively. Dotted curve is for ESC. 
}
\end{figure} 

In Fig.\ref{xsO16O16}, the calculated results of the differential cross sections 
for the $^{16}$O+$^{16}$O elastic scattering at $E/A=70$ MeV are compared with
the experimental data~\cite{Nuoffer}. 
The corresponding $^{16}$O+$^{16}$O double-folding potentials are shown
in Fig.\ref{potO16O16}.
Here, the dotted curves are obtained from ESC, and the angular distribution
deviates substantially from the data.
Solid, dashed and dot-dashed curves are for MPa, MPb and MPc, respectively.
These sets reproduce nicely the experimental data,
though the fitting by MPc seems to be slightly worse than MPa/b.
In this double-folding model analysis, most important is the 
validity of the frozen-density approximation (FDA):
Owing to the FDA, the MPP repulsion in the density region over the
normal density contributes to folding potentials.
Such an effect can be seen in Fig.\ref{potO16O16}, where the potentials
for MPa/b/c are remarkably shallower than that for ESC.
Though reduction factors are often multiplied on the imaginary parts
in the folding model analyses~\cite{FSY},
such a reduction factor is not used in the present analysis.
The necessity to include the quartic pomeron coupling has to appear in 
the difference between results for MPa and MPb, but it cannot be found 
in the present analyses for nucleus-nucleus scattering.

Recently, the detailed analysis for the above $^{16}$O+$^{16}$O
scattering has been performed using the G-matrix
double-folding potential derived from MPa with FDA~\cite{FSY14}.
Here, it has been investigated explicitly in what densities MPPs 
contribute dominantly, and found that MPP contributions
from the density region higher than the normal density
are decisively important for resultant angular distributions.
Thus, we can say that valuable information of the EoS in high-density 
region can be obtained from double-folding potentials with FDA.

As given in \cite{YFYR}, the mass-radius relations of neutron stars
are obtained by solving the TOV equation with the neutron matter EoS.
The maximum masses for MPa/b/c are $2.5M_{\odot}$, $2.2M_{\odot}$
and $2.1M_{\odot}$, respectively.

\section{$Y\!N$ interaction and hypernuclei}

Let us study here the properties of $Y\!N$ G-matrix interactions
derived from ESC in symmetric nuclear matter including a single
hyperon ($\Lambda$ or $\Sigma$).
Then, the correlations induced by baryonic coupling interactions 
such as $\Lambda\!N$-$\Sigma\!N$ ones are renormalized into 
single-channel parts of G-matrices.  
The hypernuclear phenomena and the underlying $Y\!N$ interaction models 
are linked through the models of hypernuclei and the $Y\!N$ G-matrix 
interactions, and then the hypernuclear information can be used 
to test the interaction models. Here, the G-matrix calculations
are performed in the same way as \cite{Yama10}.

Here, the most important is to test the MPP+TBA parts in channels 
including hyperons. Though MPP is defined universally in all
baryon channel, TBA is introduced phenomenologically in nucleon
channels, and not defined in $Y\!N$ channels. 
Our strategy is to determine this part so as to be consistent
with hypernuclear data. As a trial, we assume it to be the same
as for nucleon channels.

The $\Lambda\!N$ G-matrix calculations are performed for ESC and MPa/b/c.
In Table~\ref{Gmat-L1} we show the potential energies $U_\Lambda$
for a zero-momentum $\Lambda$ and their partial-wave contributions
$U_\Lambda(^{2S+1}L_J)$ at normal density $\rho_0$ ($k_F$=1.35 fm$^{-1}$),
where a statistical factor $(2J+1)$ is 
included in $U_\Lambda(^{2S+1}L_J)$. 
As shown later, the $\Lambda$-nucleus folding potentials derived from
these G-matrices lead to $\Lambda$ spectra consistent with hypernuclear data.
As for the partial wave contributions, it is important that
the odd-state contribution is weakly attractive.
In the cases of NSC97e/f models, they are strongly repulsive~\cite{NSC97}.
Such a difference becomes remarkable in high density region
relevant to $\Lambda$ mixing in neutron star matter.
The $\Lambda$ onset density is rather increased by strong odd-state 
repulsions~\cite{Vidana00}
\begin{table}[ht]
\caption{Values of $U_\Lambda$ at normal density and partial wave
contributions in $^{2S+1}L_J$ states for ESC and MPa/b/c from 
the G-matrix calculations with  CON prescriptions (in MeV).
The value specified by $D$ gives the sum of $^{2S+1}D_J$ contributions.
}
\label{Gmat-L1}
\begin{center}
 \begin{tabular}{l|ccccccc|c}
 \hline\hline
& $^1S_0$ & $^3S_1$ & $^1P_1$ & $^3P_0$ & $^3P_1$ & $^3P_2$ & $D$ 
& $U_\Lambda$  \\
 \hline
ESC    &$-$13.3& $-$26.7& 2.6 & 0.2  & 1.8 & $-$3.2 & $-$1.6 & $-$40.0 \\
MPa    &$-$13.6& $-$25.9& 3.4 & 0.4  & 2.1 & $-$1.7 & $-$2.7 & $-$38.1 \\
MPb    &$-$13.6& $-$26.0& 3.4 & 0.4  & 2.1 & $-$1.8 & $-$2.7 & $-$38.3 \\
MPc    &$-$13.4& $-$25.1& 3.2 & 0.3  & 2.0 & $-$2.1 & $-$2.4 & $-$37.4 \\
 \hline
 \end{tabular}
\end{center}
\end{table}

\begin{figure}[htb!] 
\centering 
\includegraphics[width=8cm]{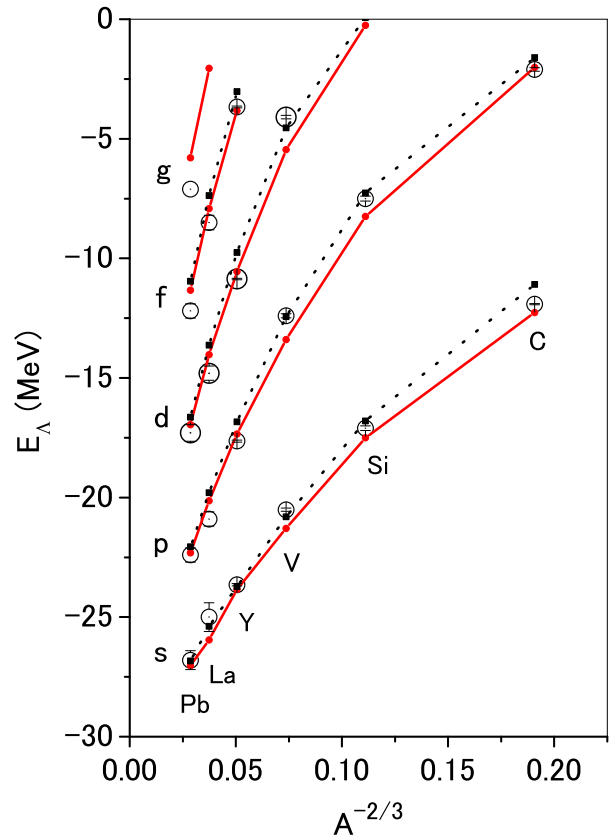} 
\caption{\small \label{spectrum}
(Color online) Energy spectra of $\Lambda$ hypernuclei
($^{13}_{\ \Lambda}$C, $^{28}_{\ \Lambda}$Si, $^{51}_{\ \Lambda}$V,
$^{139}_{\ \Lambda}$La, $^{208}_{\ \Lambda}$Pb) derived from MPa (solid lines)
and ESC (dotted lines). Experimental values are marked by open circles
}
\end{figure}

For applications to finite systems, we derive $k_F$-dependent local potentials in 
coordinate space from the G-matrices, and make $\Lambda$-nucleus folding potentials.
In this procedure, densities $\rho(r)$ and mixed densities $\rho(r,r')$ of core nuclei 
are obtained from Skyrme-HF wave functions. 
For the $k_F$-dependent parts of our localized G-matrix interactions, 
we use the averaged-density approximation:
An averaged value $\langle k_F \rangle$ is calculated 
for each $\Lambda$ state, and substituted into G-matrices.
The energy spectra of $\Lambda$ hypernuclei
($^{13}_{\ \Lambda}$C, $^{28}_{\ \Lambda}$Si, $^{51}_{\ \Lambda}$V,
$^{139}_{\ \Lambda}$La, $^{208}_{\ \Lambda}$Pb) are 
calculated with the G-matrix interactions obtained from MPa and ESC.
In Fig.\ref{spectrum}, the calculated values shown by solid (MPa) and
dashed (ESC) lines are compared with the experimental values marked 
by open circles, where the horizontal axis is given as $A^{-2/3}$.
Here, the experimental data are shifted by 0.5 MeV from the values 
given in Ref.\cite{TamHashi}, which has been recently proposed
according to the improved calibration~\cite{Gogami}.
Our G-matrix folding models turn out to reproduce 
the energy spectra of $\Lambda$ hypernuclei systematically
with no free parameter in both cases of ESC and MPa.
The results for MPb and MPc are very similar to that for MPa.
It should be noted that reasonable $\Lambda$ binding energies are 
obtained by taking the (MPP+TBA) parts equally to those in nucleon matter.

The similar results for MPa and ESC mean that
the MPP and TBA contributions are rather canceled
in evaluations of $\Lambda$ binding energies.
Here, the important point is that the results for 
(ESC+MPP+TBA) reproduce well the experimental values.
This is considered as a necessity condition which 
our MPP+TBA should satisfy at normal density region.

When the results for MPa and ESC are compared carefully,
we find are some interesting differences.
In Table~\ref{Y89L}, the calculated values of
the energy spectra of $^{89}_{\ \Lambda}$Y for MPa and ESC
are compared with the experimental values.
%
Values in parentheses are averaged values $\langle k_F \rangle$ 
which are evaluated self-consistently with solved $\Lambda$
wave functions.
The $^{89}_{\ \Lambda}$Y data has been measured with high statistics
and the obtained energy spectrum are very reliable.
In this case the result for MPa is found to be of better fitting
than that for ESC. The reason is that the stronger density-dependent
interaction of the former works more attractively in the upper states
with smaller values of $\langle k_F \rangle$.
As found in Fig.\ref{spectrum},
the same effect brings about the larger binding energies in light systems 
such as $^{13}_{\ \Lambda}$C with smaller values of $\langle k_F \rangle$, 
where the low-density contributions are dominant.

The stronger density dependence of MPa is due to the density-dependent
contributions of the (MPP+TBA) part.
It is expected that more systematical studies of
$\Lambda$ binding energies in future experiments elucidate 
the strength of the density dependence more quantitatively.

\begin{table}[ht]
\caption{Energy spectra (in MeV) of $^{89}_{\ \Lambda}$Y calculated 
with MPa and ESC in comparison with experimental values. 
Averaged values of $k_F$ (in fm$^{-1}$) are in parentheses.
}
\label{Y89L}
\begin{center}
 \begin{tabular}{l|cccc}
 \hline\hline
& $s$ & $p$ & $d$ & $f$ \\
 \hline
MPa   & $-$23.8 & $-$17.4 & $-$10.6 & $-$3.8  \\
      & (1.27)  &  (1.23) & (1.16)  & (1.08)  \\
ESC   & $-$23.7 & $-$16.8 & $-$9.8  & $-$3.0  \\
      & (1.28)  &  (1.23) & (1.17)  & (1.09)  \\
 \hline
  exp & $-23.7$ & $-17.6$ & $-10.9$ & $-3.7$ \\
 \hline
 \end{tabular}
\end{center}
\end{table}

\section{EoS and neutron stars}

\subsection{Hyperonic nuclear matter}

Let us derive here the EoS of baryonic matter
composed of nucleons ($N=n,p$) and hyperons ($Y=\Lambda, \Sigma^-$)
on the basis of the Brueckner theory.

We start from baryon single particle potentials.
From G-matrix elements in momentum space,
a single particle potential of $B$ particle in $B'$ matter is given by
\begin{eqnarray}
U_B(k)&=&\sum_{B'} U_{B}^{(B')}(k) 
\nonumber
\\
 &=& \sum_{B'} \sum_{k',k_F^{(B')}} \langle kk'|G_{BB',BB'}|kk'\rangle
\end{eqnarray}
with $B,B'=N,Y$. Here, spin isospin quantum numbers are implicit.
The energy density is given by
\begin{eqnarray}
\varepsilon&=&
\varepsilon_{kin}+\varepsilon_{pot} 
\nonumber
\\
&=& 2\sum_{B} \int_0^{k_F^B} \frac{d^3k}{(2\pi)^3}
\left\{ 
\frac{\hbar^2 k^2}{2M_B}+\frac 12 U_B(k)\right\} 
\end{eqnarray}
%
Then, we have
$$\int_0^{k_F^B} \frac{k^2 dk}{\pi^2} U_B^{(B')}(k)=
\int_0^{k_F^{B'}} \frac{k^2 dk}{\pi^2} U_{B'}^{(B)}(k)$$. 

\noindent
Considering $\rho_B=\frac{(k_F^B)^3}{3\pi^2}$
\begin{eqnarray}
\frac{\partial}{\partial \rho_B}{\cal U}_B^{(B')}=
U_B^{(B')}(k_F^B)+ \int_0^{k_F^{B}} \frac{k^2 dk}{\pi^2} 
\frac{\partial U_{B}^{(B')}(k)}{\partial \rho_B}
\end{eqnarray}
The second term leads to the rearrangement contribution.

The baryon number density is given as $\rho=\sum_B \rho_B$,
$\rho_B$ being that for component $B$.
Then, the chemical potentials $\mu_B$ and pressure $P$
are expressed as
\begin{eqnarray}
&&\mu_B = \frac{\partial \varepsilon}{\partial \rho_B} \ , 
\label{chem} \\
&& P = \rho^2 \frac{(\partial \varepsilon/\rho)}{\partial \rho_B}
 =\sum_B \mu_B \rho_B -\varepsilon \ .
\label{press}
\end{eqnarray}

\begin{figure}[htb!] 
\centering 
\includegraphics[width=8cm]{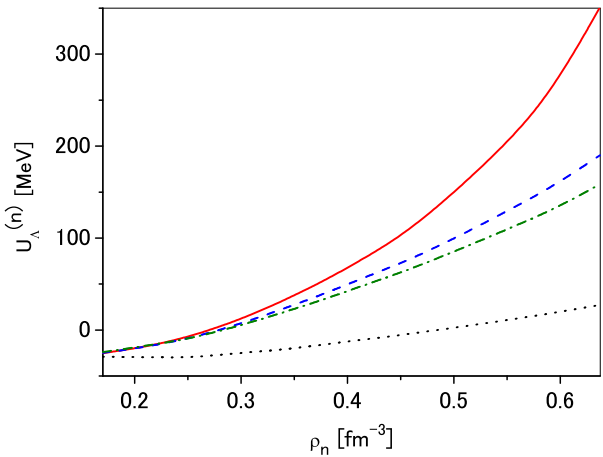} 
\caption{\small \label{ULam}
(Color online) $U_\Lambda^{(n)}$ for $\rho_\Lambda/\rho_n=0.2$ 
as a function of $\rho_n$. Solid, dashed, dot-dashed and dotted 
curves are for MPa, MPb, MPc and ESC, respectively.}
\end{figure} 

\begin{figure}[htb!] 
\centering 
\includegraphics[width=8cm]{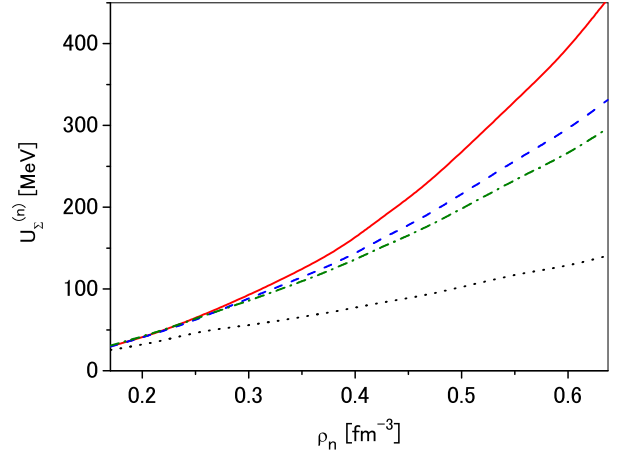} 
\caption{\small \label{USig}
(Color online) $U_{\Sigma^-}^{(n)}$ for $\rho_\Sigma/\rho_n=0.2$ 
as a function of $\rho_n$. Also see the caption of Fig.\ref{ULam}}
\end{figure} 

In Fig.\ref{ULam} and Fig.\ref{USig},
$U_\Lambda^{(n)}$ for $\rho_\Lambda/\rho_n=0.2$ and 
$U_{\Sigma^-}^{(n)}$ for $\rho_\Sigma/\rho_n=0.2$ 
are drawn as a function of $\rho_n$, respectively.
Here, solid, dashed, dot-dashed and dotted curves are
for MPa, MPb, MPc and ESC, respectively.
In the cases of MPa/b/c, we find the large repulsive contributions
from their MPP parts. The solid curves (MPa) are steeper than
the dashed (MPb) and dot-dashed (MPc) curves due to the
four-body repulsion included in MPa.
The values of $U_{\Sigma^-}^{(n)}$ should be noted to be substantially
repulsive even for ESC without MPP contributions.
The $n\Sigma^-$ interactions are dominated by contributions in 
$^3S_1$ $T=3/2$ states. The strongly repulsive contribution in this state
is due to the Pauli-forbidden state effect taken into account by 
strengthening the pomeron coupling in the ESC modeling~\cite{ESC08}.

We introduce some approximations to calculate the energy density
of baryonic matter:
(1) Hyperonic energy densities including $\Lambda$ and $\Sigma^-$
are obtained from calculations of $n+p+\Lambda$ and
$n+p+\Sigma^-$ systems, respectively.
(2) The parabolic approximation is used to treat asymmetries 
between $n$ and $p$ in $n+p$ sectors.
The calculated values of energy densities are fitted by
the following analytical parameterization~\cite{Yasutake}:
\begin{eqnarray}
&& \varepsilon_{pot}(\rho_n,\rho_p,\rho_\Lambda,\rho_\Sigma) = E_N \rho_N
\nonumber \\
&& +(E_\Lambda +E_{\Lambda \Lambda}) \rho_\Lambda
+(E_\Sigma +E_{\Sigma \Sigma}) \rho_\Sigma \ .
\label{eq:a1}
\\
&& E_z=(1-\beta)f_z^{(0)} + \beta f_z^{(1)}  
\end{eqnarray}
with $z=N$, $\Lambda$, $\Sigma$, $\Lambda \Lambda$, $\Sigma \Sigma$.
Here, we have $\beta=(1-2x_p)^2$ with $x_p=\rho_p/\rho_N$ and
$\rho_N=\rho_n+\rho_p$.
\begin{eqnarray}
&& f_N^{(i)}=a_N^{(i)} \rho_N +b_N^{(i)} \rho_N^{c_N^{(i)}}
\\
&& f_y^{(i)}=A_y^{(i)} \rho_N +B_y^{(i)} \rho_N^{c_y^{(i)}}
\\
&& A_y^{(i)}=a_{y0}^{(i)}+a_{y1}^{(i)}x_p+a_{y2}^{(i)}x_p^2
\\
&& B_y^{(i)}=b_{y0}^{(i)}+b_{y1}^{(i)}x_p+b_{y2}^{(i)}x_p^2
\label{eq:a2}
\end{eqnarray}
with $i=0, 1$ and $y=$ $\Lambda$, $\Sigma$, $\Lambda\!\Lambda$, $\Sigma\!\Sigma$. 
In the above expressions, $N$, $\Lambda$ and $\Sigma$ 
($\Lambda\!\Lambda$ and $\Sigma\!\Sigma$) denote
contributions from $N\!N$, $N\!\Lambda$ and $N\!\Sigma^-$ 
($\Lambda\!\Lambda$ and $\Sigma^-\!\Sigma^-$) interactions, respectively.

Now, the G-matrix calculations with the CON choice are performed
with ESC and MPPa/b/c sets in the density regions of
$\rho_0 < \rho_B < 4\rho_0$, and the results are fitted in
the above functional forms. The values of fitted parameters for MPa 
are listed in Table \ref{param}. 
Here, $\Lambda\!\Lambda$ $(i=0)$ parts are omitted in the Table,
because their effects are negligible in the following results.
$\Sigma^- \Sigma^-$ and $\Lambda \Sigma^-$ interactions are 
not taken into account in the present work.

\begin{table}[ht]
\caption{Parameters of energy densities for MPa given by analytical forms Eq.(10)$\sim$(15).}
\label{param}
\begin{center}
\begin{tabular}{|ccc|ccc|}
 \hline
  $a_N^{(0)}$ & $b_N^{(0)}$ & $c_N^{(0)}$ 
& $a_N^{(1)}$ & $b_N^{(1)}$ & $c_N^{(1)}$ \\
$-234.8$ & 643.8 & 1.86 & 66.41 & 490.1 & 2.40 \\
 \hline
\end{tabular}
\vskip 0.3cm
\begin{tabular}{|ccc|ccc|c|}
 \hline
  $a_{\Lambda 0}^{(0)}$ & $a_{\Lambda 1}^{(0)}$ & $a_{\Lambda 2}^{(0)}$ 
& $b_{\Lambda 0}^{(0)}$ & $b_{\Lambda 1}^{(0)}$ & $b_{\Lambda 2}^{(0)}$ & $c_\Lambda^{(0)}$ \\
$-436.4$ & 1198. & $-2790.$ & 1648. & $-3970.$ & 12730. & 2.29 \\
  $a_{\Lambda 0}^{(1)}$ & $a_{\Lambda 1}^{(1)}$ & $a_{\Lambda 2}^{(1)}$ 
& $b_{\Lambda 0}^{(1)}$ & $b_{\Lambda 1}^{(1)}$ & $b_{\Lambda 2}^{(1)}$ & $c_\Lambda^{(1)}$ \\
$-215.2$ & 14.17 & $-202.7$ & 1117. &  2283. & $-198.5$ & 2.56 \\
 \hline
  $a_{\Sigma 0}^{(0)}$ & $a_{\Sigma 1}^{(0)}$ & $a_{\Sigma 2}^{(0)}$ 
& $b_{\Sigma 0}^{(0)}$ & $b_{\Sigma 1}^{(0)}$ & $b_{\Sigma 2}^{(0)}$ & $c_\Sigma^{(0)}$ \\
$-5.017$ & $-618.6$ & 1444. &  382.1 & 1803. & $-2748.$ & 2.00 \\
  $a_{\Sigma 0}^{(1)}$ & $a_{\Sigma 1}^{(1)}$ & $a_{\Sigma 2}^{(1)}$ 
& $b_{\Sigma 0}^{(1)}$ & $b_{\Sigma 1}^{(1)}$ & $b_{\Sigma 2}^{(1)}$ & $c_\Sigma^{(1)}$ \\
 100.4 & 178.5 & $-186.2$ & 909.9 & 1875. & $-1071.$ & 2.65 \\
 \hline
  $a_{\Lambda\Lambda 0}^{(1)}$ & $a_{\Lambda\Lambda 1}^{(1)}$ & $a_{\Lambda\Lambda 2}^{(1)}$ 
& $b_{\Lambda\Lambda 0}^{(1)}$ & $b_{\Lambda\Lambda 1}^{(1)}$ & $b_{\Lambda\Lambda 2}^{(1)}$ 
& $c_{\Lambda\Lambda}^{(1)}$ \\
 .7345 & $-92.92$ & 57.59 & $-4.144$ & 419.9 & 480.7 & 2.27 \\
 \hline
\end{tabular}
\end{center}
\end{table}

\subsection{EoS of hyperon-mixed neutron-star matter}

Our neutron-star matter is composed of $n$, $p$, $e^-$, $\mu^-$, 
$\Lambda$ and $\Sigma^-$.
The equilibrium conditions are summarized as follows:

\noindent
(1) chemical equilibrium conditions,
\begin{eqnarray}
\label{eq:c1}
&& \mu_n = \mu_p+\mu_e \\
&& \mu_\mu = \mu_e \\
&& \mu_\Lambda = \mu_n \\
&& \mu_{\Sigma^-} =\mu_n + \mu_e
\label{eq:c2}
\end{eqnarray}
\noindent
(2) charge neutrality,
\begin{eqnarray}
\rho_p = \rho_e +\rho_\mu + \rho_{\Sigma^-}
\end{eqnarray}
\noindent
(3) baryon number conservation,
\begin{eqnarray}
\rho = \rho_n +\rho_p +\rho_\Lambda + \rho_{\Sigma^-}
\label{eq:c3}
\end{eqnarray}

When the analytical expressions (\ref{eq:a1})$\sim$(\ref{eq:a2})
are substituted into the chemical potentials (\ref{chem}),
the chemical equilibrium conditions (\ref{eq:c1})$\sim$(\ref{eq:c2})
are represented as equations for densities $\rho_a$
($a=$ $n$, $p$, $e^-$, $\mu^-$, $\Lambda$ and $\Sigma^-$).
Then, equations (\ref{eq:c1})$\sim$(\ref{eq:c3})
can be solved iteratively.

\begin{figure}[htb!] 
\centering 
\includegraphics[width=8cm]{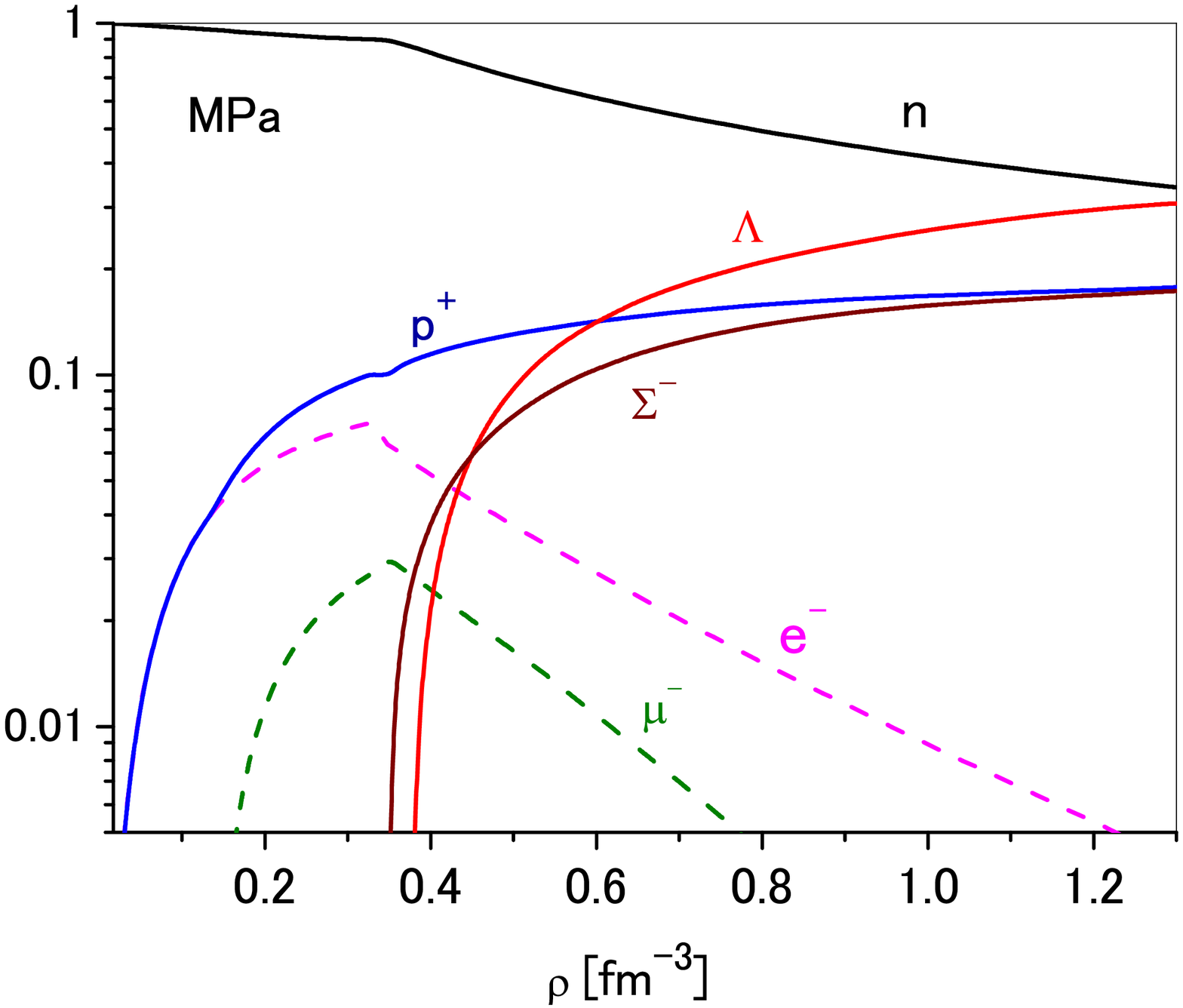} 
\caption{\small \label{chemi1}
(Color online) Composition of hyperonic neutron-star matter 
in the case of MPa.
}
\end{figure} 

\begin{figure}[htb!] 
\centering 
\includegraphics[width=8cm]{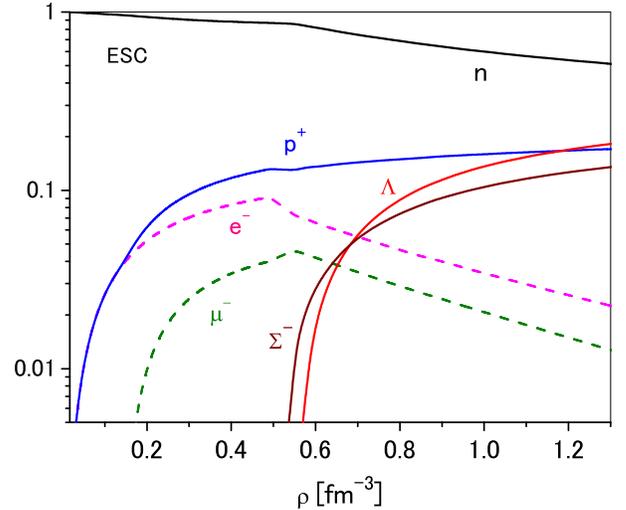} 
\caption{\small \label{chemi2}
(Color online) Composition of hyperonic neutron-star matter 
in the case of ESC.
}
\end{figure} 

In Fig.~\ref{chemi1} and Fig.~\ref{chemi2}, the matter compositions 
are shown in the cases of MPa and ESC, respectively. 
Comparing the two figures, we note some effects of
MPP contributions: (1) The onset densities of hyperon mixing
for MPa are lower than those for ESC, (2) Hyperon components for
MPa are larger than those for ESC, (3) Larger hyperon components
for MPa are covered by smaller components of $n$, $e^-$ and $\mu^-$,
and proton components are not so different from each other.
In Table~\ref{Onset}, the onset densities of hyperon mixing
are given for MPa/b/c and ESC.
Thus, increasing of MPP repulsions are found to enhance
hyperon mixings. 

\begin{table}[ht]
\begin{center}
\caption{Onset densities in fm$^{-3}$.
}
\vskip 0.2cm
\begin{tabular}{c|cc}
\hline\hline
 Model & $\Sigma^-$ & $\Lambda$  \\
\hline
  MPa   & 0.34 & 0.36  \\
  MPb   & 0.37 & 0.42  \\
  MPc   & 0.39 & 0.45  \\
  ESC   & 0.52 & 0.54  \\
\hline
\end{tabular}
\label{Onset}
\end{center}
\end{table}

Here, let us see the role of MPP repulsions more in detail. 
The repulsions among neutrons make single particle potentials 
shallower, which allows more easily conversions of neutrons 
into hyperons. These effects are partially canceled out by 
the repulsions including hyperons, which make shallower 
hyperon single particle potentials. 
As seen in Table \ref{Onset}, for instance, 
the onset densities of $\Sigma^-$
and $\Lambda$ are 0.52 and 0.54 fm$^{-3}$ for ESC, respectively. 
If only the (MPP+TBA) contributions among nucleons are taken 
into account in the case of MPa, both of them are 0.32 fm$^{-3}$.
Then, the values of 0.34 and 0.36 fm$^{-3}$ for MPa are understood 
as a result of partial cancelling of the MPP repulsive effects.

Pressures (\ref{press}) are derived from 
determined values of densities and chemical potentials. 
In Fig.~\ref{press}, the calculated values of pressure $P$ are drawn as
a function of baryon density $\rho$ in the cases of MPa (upper curves) 
and ESC (lower curves), where solid and dotted curves are with and 
without hyperon mixing, respectively.  
The dashed curve is with hyperon mixing, where the MPP+TBA parts are
included only in nucleon channels.
The difference between the two dotted curves are due to the MPP repulsive
contributions among nucleons, and the remarkable softening from the upper 
dotted curve to the dashed curve is brought about by hyperon mixing.
This softening is substantially recovered when the MPP contributions are 
included universally among baryons.

\begin{figure}[htb!] 
\centering 
\includegraphics[width=8cm]{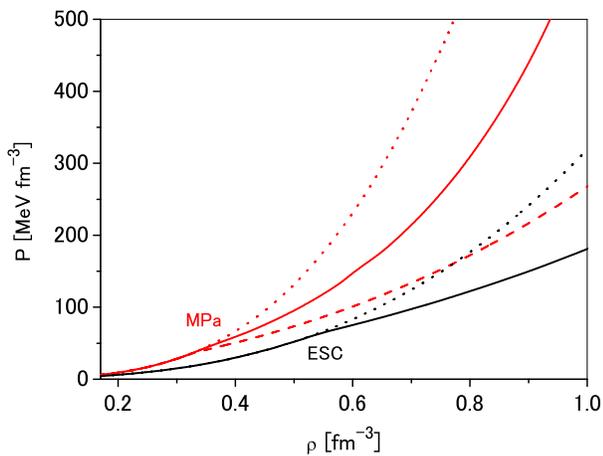} 
\caption{\small \label{press}
(Color online) Pressure $P$ as a function of baryon density $\rho$ 
in the cases of MPa (upper curves) and ESC (lower curves).
Solid and dotted curves are with and without hyperon mixing, respectively.
In dashed curve with hyperon mixing, MPP+TBA parts are switched off
in channels including hyperons.
}
\end{figure}

\subsection{Neutron stars}

Using the EoS of hyperonic neuron matter, we solve the 
TOV equation for the hydrostatic structure
of a spherical non-rotating star, and obtain the mass and 
radius of neutron stars. 
The EoS's for MPa/b/c and ESC are used $\rho > \rho_0$
Below $\rho_0$ we use the EoS of the crust obtained
in \cite{Baym1,Baym2}. Then, the EoS's for $\rho > \rho_0$
and $\rho < \rho_0$ are connected smoothly.

\begin{figure}[ht] 
\begin{center} 
\includegraphics*[width=8cm]{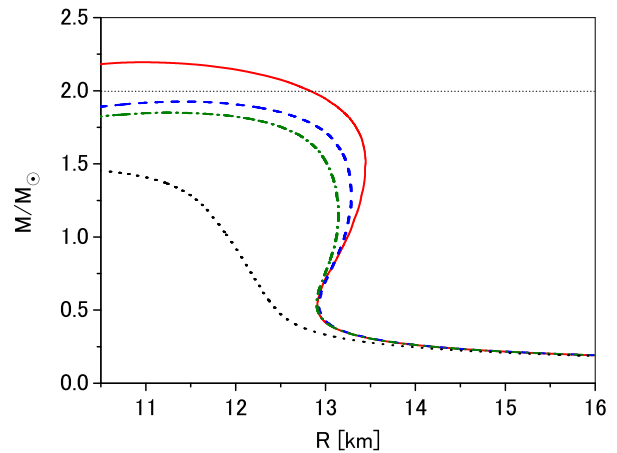} 
\caption{\small \label{hypstar1}
(Color online) Neutron-star masses as a function of the radius $R$.
Solid, dashed, dot-dashed and dotted curves are for MPa/b/c 
and ESC, respectively.}
\end{center}
\end{figure} 

In Fig.~\ref{hypstar1}, neutron-star masses are drawn as 
a function of radius, where solid, dashed, dot-dashed 
and dotted curves are for MPa/b/c and ESC, respectively. 
Then, calculated values of maximum masses $M/M_{\odot}$ 
are 2.20$M_{\odot}$, 1.93$M_{\odot}$ and 1.85$M_{\odot}$ 
for MPa/b/c, respectively. These values are smaller by 
about 0.3$M_{\odot}$ than the values without hyperon mixing.
Thus, the maximum mass only for MPa is noted to be 
substantially larger than the observed value of 
$\sim 2M_{\odot}$ owing its four-body repulsive contribution.
It should be noted that the difference between MPa and MPc
comes from the four-body repulsion included in the former,
because MPc is made by switching off the four-body part
from MPa. On the other hand, MPb is designed so as to 
reproduce the repulsive effect of MPa in the 
$^{16}$O-$^{16}$O scattering without the four-body
repulsive part.
The difference between MPa and MPb is originated from
the steeper EoS of MPa by the four-body repulsion
in the high-density region.

\begin{figure}[ht] 
\begin{center} 
\includegraphics*[width=8cm]{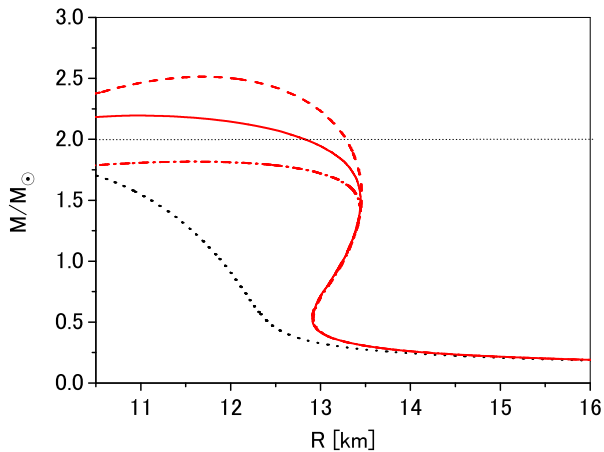}
\caption{\small \label{hypstar2}
(Color online) Neutron-star masses as a function of the radius $R$.
Dashed and dotted curves are obtained from
MPa and ESC, respectively, without hyperon mixing.
Solid (dot-dashed) curve is obtained with hyperon mixing, 
where MPP contributions are included 
in all baryons universally (only in nucleon sectors).
}
\end{center}
\end{figure} 

\begin{figure}[ht] 
\begin{center} 
\includegraphics*[width=8cm]{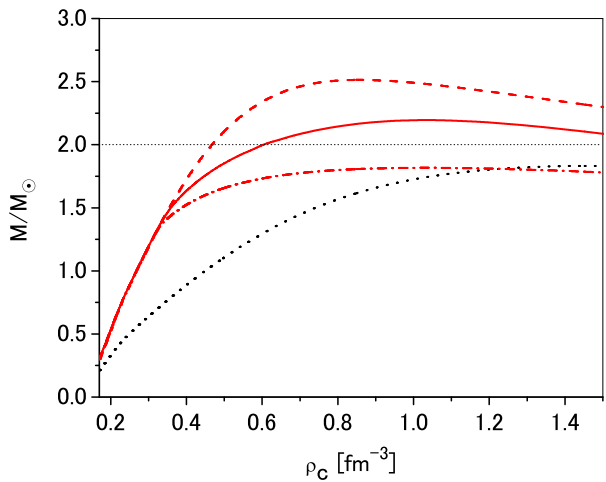}
\caption{\small \label{hypstar3}
(Color online) Neutron-star masses as a function of the central density $\rho_c$.
Also see the caption of Fig.\ref{hypstar2}
}
\end{center}
\end{figure}

The M-R relations in Fig.~\ref{hypstar2} demonstrate the 
effects of hyperon mixings and MPP contributions.
Here, dotted and dashed curves are obtained from
ESC and MPa, respectively, without hyperon mixing.
Then, the remarkable difference between dotted and dashed 
curves are owing to the MPP contributions among nucleons
included in MPa.
Solid and dot-dashed curves are obtained with hyperon mixing. 
Here, the MPP contributions in the former (latter) are included 
in all baryons universally (only in nucleon sectors).
When the MPP contributions are included only in nucleon sectors,
the maximum mass in the dashed curve with 2.51$M_{\odot}$
is strongly reduced to 1.82$M_{\odot}$ in the dot-dashed one 
by the effect of hyperon mixing.
When the MPP contributions are taken into account universally
in all baryons, the maximum mass is recovered to 2.20$M_{\odot}$
in the solid curve.
We can find the same demonstration in Fig.~\ref{hypstar3}
where the corresponding curves of neutron-star masses are
drawn as a function of central density $\rho_c$.

\section{Conclusion}

The existence of neutron stars with $2M_{\odot}$ give a severe 
condition for the stiffness of EoS of neutron-star matter, 
namely the necessity of the strong TNR.
On the other hand, the hyperon mixing in neutron-star matter 
brings about the remarkable softening of the EoS, which 
cancels the TNR effect for the maximum mass. 
As a possibility to avoid this serious problem, we introduce
the TNR-like repulsions working universally for $Y\!N\!N$, $Y\!Y\!N$ 
$Y\!Y\!Y$ as well as for $N\!N\!N$ \cite{NYT}.

On the basis of the $BB$ interaction model ESC,
we introduce the universal three-body repulsion MPP among three baryons.
The strengths of MPP are determined by fitting the observed angular 
distribution of $^{16}$O+$^{16}$O elastic scattering at $E_{in}/A=70$ MeV
with use of the G-matrix folding potential. Then,
TNA is added to MPP phenomenologically so as to reproduce the minimum 
value $\sim -16$ MeV of the energy per nucleon at normal density 
0.16 fm$^{-3}$ in symmetric nuclear matter as well as 
the $^{16}$O+$^{16}$O data. In this modeling,
the empirical values of $K$, $E_{sym}$ 
and $L$ are reproduced reasonably.
The EoS of neutron-star matter obtained from ESC+MPP+TNA is stiff 
enough to give the large neutron-star mass over $2M_{\odot}$,
when the hyperon mixing is not taken into account.

In order to study the effect of hyperon mixing to the EoS and 
mass-radius relations of neutron stars, we need to use
reliable interactions in channels including hyperons. The reliability 
of ESC in these channels have been confirmed by successful applications 
to hypernuclear systems.
Our MPP contributions exist universally in every baryonic system.
Assuming that the remaining part TNA also contributes universally as TBA,
ESC+MPP+TBA can be tested in applications to hypernuclei:
The energy spectra of $\Lambda$ hypernuclei are nicely reproduced
by the derived G-matrix interactions with no modification for TBA.
Then, it is suggested that inclusion of MPP+TBA leads to even better 
fitting than the case of using ESC part only.

The EoS of hyperonic nuclear matter is obtained from ESC+MPP+TBA
on the basis of the G-matrix approach, and the mass-radius relations
of neutron stars are derived by solving the TOV equation.
In spite of remarkable softening of EoS caused by hyperon mixing,
its substantial part is recovered owing to the MPP contributions.
As a result, the universal MPP repulsions are shown to bring about
hyperon-mixed neutron stars with masses $\sim 2M_{\odot}$.
It should be noted that our conclusion for neutron stars is
obtained essentially on the basis of terrestrial experiments
without using ad hoc parameters to stiffen the EoS.


\end{document}